# Modulated-bath AC calorimetry using modified commercial Peltier-elements

Rolf Lortz, Satoko Abe, Yuxing Wang, Frédéric Bouquet†, Ulrich Tutsch, and Alain Junod

Département de Physique de la Matière Condensée, Université de Genève, 24 Quai Ernest-Ansermet, CH-1211 Geneva 4, Switzerland.

We developed a new type of AC microcalorimeter based on a modulated-bath technique for measuring the specific heat of small microgram samples in the temperature range from 30-300 K, and tested it in magnetic fields up to 14 T. The device is built from a modified commercial Peltier element. The temperature of its top plate can be modulated periodically by Peltier effect, so that the oscillation is symmetrical about the temperature of the main bath. This avoids the problem of DC offsets which plagues conventional AC calorimeters. The sample is attached to a thin thermocouple cross, acting as a weak thermal link to a platform. The absence of a heater reduces the background heat capacity ('addenda') to a minimum. As an illustrative example of the performance of our device, the specific heat in fields up to 14 T of a small single crystal of the high-temperature superconductor $Bi_{2.12}Sr_{1.71}Ca_{1.22}Cu_{1.95}O_y$ is determined.

† Present address: Laboratoire de Physique des Solides, CNRS Université Paris-Sud, 91405 Orsay Cedex, (France).

## I. INTRODUCTION

As new materials are often available only in small quantities, the development of devices allowing high-resolution measurements of thermal properties of tiny samples with masses of the order of several micrograms down to a few hundreds of nanograms is of particular interest. The temperature dependence of the specific heat is one of the most basic and informative properties, however calorimetric measurements of samples with such tiny masses require special techniques[1-8]. In this article we present a new modulated-bath AC calorimeter for samples with masses of a few micrograms built from a commercial Peltier element, which presents the advantage of keeping the average temperature of the sample close to that of the thermal bath.

## II. METHOD

Besides standard, adiabatic methods where the sample is thermally isolated from its surroundings, dynamical methods with a sample of heat capacity $C$ being connected to a thermal bath through a weak thermal link of conductance $k$ appear more promising for the measurement of samples with small masses. Alternating current (AC) methods working at a frequency $\omega >> \tau^{-1} = k/C$ have the advantage that a thermocouple can be used as a thermal link and thermometer at the same time[1,2,9], which strongly reduces the addenda. In addition AC calorimetry allows signal averaging over many heating cycles. This leads to a high resolution for even the tiniest samples. Modulated-bath AC calorimeters[1,2] using a sine wave heating function have been proven to be extremely useful for this purpose. The sample is attached to a thermocouple cross made of thin wires, while the heater is spatially separated from the sample. In this method the heat flows through the thermocouple legs towards the sample and may be more quantitatively controlled than e.g. with optical heating[6-7,9]. Due to the absence of a sample platform the addenda are reduced to a minimum. Heat can be supplied to the sample by an electrically insulated heater located on the legs of the thermocouples measuring the temperature difference between sample and the thermal bath. The use of such a Joule heater as a source of the temperature modulation has the disadvantage that the heat is produced with the square of the current, and therefore always flows in the same direction. This causes a DC offset which plagues conventional AC calorimeters, since the condition $\omega\tau >> 1$ also means that the AC component of the thermocouple voltage carrying the useful information is smaller than the DC offset. Alternatively, the temperature modulation can be created by Peltier effect[3]. In the latter case the heating power is directly proportional to the alternating current, thus modulating the sample temperature symmetrically about the main bath temperature.

## III. CONSTRUCTION OF THE CALORIMETER

The basic building block of the calorimeter is a commercially produced element, employing the Peltier effect for cooling electronic devices[10]. The same elements were used as described in an earlier paper[11], where they served as sensitive heat-flow meters to measure the specific heat and the isothermal magnetocaloric effect of large samples. Here we use them in an inverse way by applying the Peltier effect to modulate the sample temperature to extract the specific heat using an AC technique.

For the construction of the calorimeter the base plate of the Peltier element is soldered with In-Sb eutectic (melting temperature 117°C) onto a copper platform, representing a main thermal bath at a temperature $T_B$. To avoid damage to the Peltier elements due to differential thermal expansion, a loop of thin (50-100 μm) copper wire is embedded in the solder and acts as a buffer between the base plate of the Peltier element and the copper platform. The main bath temperature is controlled using the temperature regulation of a variable temperature insert in a magnet cryostat with a field up to 16 T. The calorimeter is continuously evacuated by a turbomolecular pump to reduce convective heat losses and to avoid anomalies due to the condensation/evaporation of residual gases.

The top plate of the Peltier element (3.2 x 3.2 mm or 6.4 x 6.4 mm, see Fig. 1a for a schematic diagram of the element) can be considered as a modulated bath. The temperature modulation is driven by the output of a lock-in amplifier connected to the terminals of the device. The top plate of the element is made of alumina, coated with a thin layer of copper. In the middle of the top plate a cavity of approximately 2 mm diameter and ~0.3 mm depth has been ground with a handheld diamond grinder. The remaining copper coating has been cut with a diamond saw, so that four electrically insulated copper pads in the corners and one additional pad on one side of the top plate are left (Fig. 1b). A Chromel-Constantan thermocouple ($th_1$ in Fig. 1b), made of spot-welded 25 $\mu$m ∅ wire, is either soldered with indium or glued with silver epoxy[12] to one pad. It monitors the oscillating part of the temperature difference between the top plate and the main bath ($\Delta T_{P\text{-}B} = S\, U_{P\text{-}B}$ where $S$ is the Seebeck coefficient of standard type E Chromel-Constantan thermocouples and $U_{P\text{-}B}$ is the voltage measured between its legs where they are thermally anchored to a common temperature). A thermocouple cross ($th_2$ in Fig. 1b) – made of the same material as $th_1$, but flattened after spotwelding – is thermally anchored in the same way to the four remaining pads. Two of the legs are continued by a pair of phosphor-bronze wires of the same diameter. They measure the difference between the sample temperature and the modulated-bath temperature ($\Delta T_{S\text{-}P} = S\, U_{S\text{-}P}$). The other two legs are continued using the same thermocouple material to heat-sink chips located on the main bath, so that they measure the difference between the oscillating sample temperature and the (constant) main bath temperature ($\Delta T_{S\text{-}B} = S\, U_{S\text{-}B}$).

The sample is glued onto the thermocouple cross using a mixture of 95 % ethanol and 5 % GE7031 varnish[13]. To do so, a drop is deposited into the hole on the top plate using a syringe so that the liquid fills the space under the thermocouple cross. The surface tension of the liquid helps to gently push the sample onto the thermocouple cross, where it remains stuck after the varnish has dried. To remove the sample it is sufficient to put one or two drops of ethanol into the hole to wash away the varnish. To facilitate the determination of the contribution of the glue to the total heat capacity (see section V), care should be taken to keep the same composition in all experiments. GE7031 is a well-known thermal compound in specific-heat experiments. A characterization of its thermal properties can be found in Ref.[14].

For the specific-heat measurement an AC voltage signal of typically $U_{ac}$ = 0.5-2 V (effective value) is set at the output of the lock-in amplifier connected to the terminals of the Peltier element (note that the current is determined essentially by the output resistance of the current source, 50 Ω, and that of the connecting wires, about 30 Ω; the resistance of the Peltier element itself is only 1 Ω). The temperature of the main bath $T_B$ is measured using a Cernox sensor[15], mounted just besides the Peltier element on the copper block. $T_B$ is ramped up or down at a rate of 0.5 – 1 K/min, using the temperature sweep facility of the variable temperature insert of the cryostat. The signals of the thermocouples ($\Delta T_{P\text{-}B}$, $\Delta T_{S\text{-}B}$ and $\Delta T_{S\text{-}P}$) are amplified $10^4$ times by picovolt DC amplifiers[16], when the frequency is low enough (~ <1 Hz), and fed into digital lock-in amplifiers[17].

## IV. PERFORMANCE

The configuration of the thermal anchoring of the four legs of the thermocouples allows us to measure the temperature of the sample either relative to the modulated bath (through $U_{S\text{-}P}$) or relative to the main bath (through $U_{S\text{-}B}$). We shall make a distinction between the specific-heat techniques associated with the use of either one of these signals. In any case $U_{P\text{-}B}$ also needs to be measured since it represents the thermal excitation provided to the sample. Fig. 2 shows a schematic diagram of the thermal configuration. $T_S$, $T_P$ and $T_B$ are the temperature of the sample, the platform and the thermal bath; $k$ is the heat leak between the sample and the platform, and $C$ is the heat capacity of the sample (including contributions from the addenda). $k$ is essentially due to the thermal conduction along the thermocouple. The system obeys the equation:

$$k(T_S - T_P) + C\frac{dT_S}{dt} = 0 \qquad (1)$$

As usually $T_S$ conveys the information on the specific heat. The thermocouples allow us to measure either $T_S - T_B \equiv \Delta T_{S\text{-}B}$ (with $U_{S\text{-}B}$), or $T_S - T_P \equiv \Delta T_{S\text{-}P}$ (with $U_{S\text{-}P}$; see Fig. 1 b). Both can be used to determine $C$, but the frequency dependence of the signals is different. In the following, we further define $T_P - T_B = \Delta T_{P\text{-}B}$, and adopt implicitly the complex notation.

**A) Modulated-bath AC calorimetry, using $\Delta T_{S\text{-}P}$:**

When $\Delta T_{S\text{-}P}$ is used to measure the sample temperature, the device is used in the modulated-bath AC calorimetric mode. The peculiarity of our system is that the Joule heater is replaced by a Peltier element with the advantages already mentioned. Since $T_B$ is kept constant or slowly swept during the experiment, $dT_B/dt$ may be neglected. Thus $dT_S/dt = d(\Delta T_{S\text{-}P} + \Delta T_{P\text{-}B})/dt$. Eq. (1) can then be rewritten as:

$$\tau\left(\frac{d(\Delta T_{S-P})}{dt} + \frac{d(\Delta T_{P-B})}{dt}\right) + \Delta T_{S-P} = 0 \qquad (2)$$

where $\tau = C/k$ is the time constant. The harmonic solution is:

$$\Delta T_{P-B} = \left|\Delta T_{P-B}\right| e^{i\omega t}$$

and

$$\Delta T_{S-P} = \left|\Delta T_{S-P}\right| e^{i(\omega t + \varphi)} \qquad (3)$$

with $\omega$ the excitation frequency. The phase shift $\varphi$ and the amplitudes satisfy:

$$\tan\varphi = \frac{1}{\omega\tau} \qquad (4)$$

$$\left(\frac{\left|\Delta T_{S-P}\right|}{\left|\Delta T_{P-B}\right|}\right)^2 \left[1 + \frac{1}{(\omega\tau)^2}\right] = 1 \qquad (5)$$

The specific heat can be deduced either from the phase:

$$C = \frac{k}{\omega}\tan^{-1}\varphi \qquad (6)$$

or from the amplitude ratio:

$$C = \frac{k}{\omega} \sqrt{\frac{1}{\frac{|\Delta T_{P-B}|^2}{|\Delta T_{S-P}|^2} - 1}} \qquad (7)$$

Below the cutoff frequency $\omega_0 = 1/\tau$ the sample temperature $\Delta T_{S-B}$ asymptotically approaches the temperature modulation of the Peltier element $\Delta T_{P-B}$ and thus the signal on the thermocouple $\Delta T_{S-P}$ is small. This frequency range is favorable to extract the heat capacity using Eq. (6) or (7) from the signal at the excitation frequency and not the second harmonic as would be the case for Joule heating. This method allows us thus to measure the specific heat at a fixed working frequency $\omega < \omega_0$, whereas in standard AC calorimetry the specific heat enters the signal mainly above the cutoff frequency. We note that the condition $\omega < \omega_0$ is not a necessary one, but it leads to a better resolution because $\frac{|\Delta T_{P-B}|^2}{|\Delta T_{S-P}|^2} >> 1$ in Eq. (7). Working at a frequency $\omega < \omega_0$ may be of advantage in the presence of additional time constants due to a low thermal conductivity of the sample or imperfect thermal coupling between the sample and the thermocouple. In addition, the low frequency range leaves the choice between DC preamplifiers and transformers open to optimize the signal / noise ratio. The drawback is that the sample temperature is modulated with a large amplitude, close to that of the modulated bath, while the useful signal $|\Delta T_{S-P}|$ is only a fraction of that. This may broaden specific-heat anomalies at phase transitions. The working frequency during a measurement should thus be chosen rather close to the cutoff frequency where $|\Delta T_{S-P}|$ is of the same order of magnitude as $|\Delta T_{S-B}|$.

**B) Standard AC calorimetry, using $\Delta T_{S-B}$:**

When $\Delta T_{S-B}$ is used to measure the sample temperature, the frequency dependence of the sensitivity is different, and looks more like that of a conventional AC calorimeter[18]. The only difference is that the source of the temperature modulation is not located on the sample but on the legs of the thermocouples. Eq. (1) can be rewritten as:

$$\Delta T_{S-B} - \Delta T_{P-B} + \tau \frac{d(\Delta T_{S-B})}{dt} = 0 \qquad (8)$$

with the harmonic solution

$$\Delta T_{P-B} = |\Delta T_{P-B}| e^{i\omega t}$$

and

$$\Delta T_{S-B} = |\Delta T_{S-B}| e^{i(\omega t + \psi)} \qquad (9)$$

where the phase shift $\psi$ and the amplitudes satisfy:

$$\tan\psi = -\omega\tau \qquad (10)$$

$$\left(\frac{|\Delta T_{S-B}|}{|\Delta T_{P-B}|}\right)^2 \left[1 + (\omega\tau)^2\right] = 1 \qquad (11)$$

The specific heat can be deduced either from the phase:

$$C = -\frac{k}{\omega} \tan\psi \qquad (12)$$

or from the amplitude:

$$C = \frac{k}{\omega}\sqrt{\frac{\left|\Delta T_{P-B}\right|^2}{\left|\Delta T_{S-B}\right|^2} - 1} \tag{13}$$

The signal $|\Delta T_{S-B}|$ is maximal at low frequencies where it asymptotically approaches $|\Delta T_{P-B}|$, whereas it decreases above the cutoff frequency. Here, it is more favorable to measure the heat capacity at a fixed working frequency $\omega > \omega_0$. In this case the condition $\frac{\left|\Delta T_{P-B}\right|^2}{\left|\Delta T_{S-B}\right|^2} >> 1$ in Eq. (13) ensures a better resolution on $C$. The advantage of this method is that the temperature modulation of the sample can be measured at relatively high frequency, allowing a faster sweep rate $dT_B/dt$, while still averaging over many cycles with the lock-in technique. But most important, the measured signal $|\Delta T_{S-B}|$ represents directly the real amplitude of the modulation of the sample temperature. Contrary to the previous method, it can be kept at a sufficiently low level to avoid broadening effects near phase transitions.

Fig. 3a shows the frequency dependence of the amplitude of $\Delta T_{S-P}$, $\Delta T_{S-B}$ and $\Delta T_{P-B}$ for a sample with a mass of 26 μg at $T$=90 K. The response of the modulated bath is shown by $|\Delta T_{P-B}|$. It follows the behavior of a system which can be described by Eq. (14) with $C_P$ the heat capacity of the top plate of the Peltier element, $k_P$ the heat conductance of the thermopile between the top and base plates, and $\tau_P=C_P/k_P$:

$$\left|\Delta T_{P-B}\right| = T_0\sqrt{\frac{1}{1+\omega^2\tau_P^2}} \tag{14}$$

By fitting this formula to the data, we find $\tau_P$=0.32 s at 90 K for the time constant of the Peltier element used to create the temperature modulation. Fig. 3b shows the frequency dependence of $|\Delta T_{S-P}|/|\Delta T_{P-B}|$ and $|\Delta T_{S-B}|/|\Delta T_{P-B}|$ together with a fit using Eq. (5) and (11), respectively. The data were measured here without preamplifiers, to avoid introducing additional poles in the frequency dependence. The theoretical curves are in good agreement with the data in the relevant frequency range below ~2 Hz. The frequency corresponding to the crossing point of the two fitting functions determines the main time constant (in our case for a 26 μg $Bi_{2.12}Sr_{1.71}Ca_{1.22}Cu_{1.95}O_y$ sample at 90 K: $\tau$ =0.88 s, slightly above that of the Peltier element). This ensures that sufficient power is available to create a temperature modulation of convenient amplitude over the useful range of working frequencies (~0.05 to ~3 Hz). The frequency dependence of the thermocouple response should be tested before each experiment at a few different temperatures to determine the ideal working frequency which fulfills the condition $\omega < \omega_0$ ($\Delta T_{S-B}$) or $\omega > \omega_0$ ($\Delta T_{S-P}$) over the entire temperature range. In some cases we found deviations from the theoretical curves at higher frequencies, indicating a bad thermal coupling between the sample and the thermocouple cross. In such a case the sample had to be mounted again.

The upturn in the modulation above 4 Hz in Fig. 3a is associated with the internal time constant of the Bi-Te material in the Peltier cell. This happens more than one decade above the typical cutoff frequency of the sample / thermocouple system, so that this effect is not a real limitation. Fig. 3c shows the amplitudes of the responses of $\Delta T_{S-P}$, $\Delta T_{S-B}$ and $\Delta T_{P-B}$ as a function of amplitude of the driving AC voltage ($U_{ac}$, as set on the output display of the lock-in amplifier).

The dependence is linear, as expected for reversible Peltier heating $P(t) = \pi I(t)$, where $\pi$ is the Peltier coefficient and $I(t)$ the driving current. To test if there is any significant component due to Joule heating $P(t) = RI^2(t)$ from the $R \sim 1\ \Omega$ internal resistance of the Peltier element, we also measured the second harmonic of $\Delta T_{P\text{-}B}$ and $\Delta T_{S\text{-}B}$. The signal at $T$=100 K, using an AC voltage drive of $U_{ac}$=1.8 V (effective value, i.e. about 20 mA) and a frequency of 0.555 Hz, was smaller than 5 mK, nearly two orders of magnitude smaller than the signal at the driving frequency. The typical DC offset found between the temperature of the sample and that of the main bath is smaller than 0.1 K at all temperatures, as long as the temperature sweep rate during the experiment is kept below 1 K/min. Therefore the DC voltage does not saturate the input of the amplifiers, and the average temperature of the sample is essentially that of the main bath.

The design of the calorimeter thus allows us to extract the heat capacity by measuring the sample temperature $T_S$ with two different thermocouple configurations having a different frequency dependence. Depending on the value of the cutoff frequency (directly related to the heat capacity of the sample and the heat conductance of the thermocouple), the most suitable frequency range can be chosen. This may be important as high sensitivity DC preamplifiers can only be used at low frequencies (typically < 1 Hz) whereas transformers only work at higher frequencies.

Finally note that the amplitudes enter Eq. (7) and (13) only as ratios. This favors the determination of $C$ based on the amplitudes rather than the phases, since the frequency response of identical amplifiers in both channels is compensated. the $U_{P\text{-}B}$ channel does not convey any information on the heat capacity and may be smoothed to improve the signal / noise ratio.

## V. CALIBRATION OF THE CALORIMETER

In order to calculate the absolute value of the specific heat, the heat conductance $k$ of the thermocouple leads and the heat capacity of the addenda, i.e. part of the thermocouple cross glued to the sample and varnish, have to be determined independently. A first estimation for $k(T)$ of a Chromel/Constantan thermocouple can be obtained from literature data[1]. A precise method to estimate the relevant length of the thermocouple legs is to measure them on an enlarged photograph of the calorimeter with the sample mounted. We use the lengths from the edge of the sample, covering the thermocouple cross, to the top plate of the Peltier element.

In order to determine $k(T)$ and the contribution from the addenda more accurately, we use two silver samples of high purity with different masses (85 μg and 48 μg). The contribution of the thermocouple cross to the total heat capacity depends somewhat on the shape and size of the sample. The associated uncertainties can be minimized if two samples with a flat surface of the same size are chosen. More generally we use samples with a flat surface comparable to our calibration samples (typically 0.5 x 0.5 mm$^2$) when absolute accuracy is needed. Assuming that the addenda heat capacity remains unchanged in the two Ag calibration runs, it cancels out in the difference. This difference, which corresponds to a bare 37 μg sample, can be used to extract $k(T)$ based on reference data[19]. We believe that with this method an accuracy of 2 % can be obtained in the determination of $k(T)$. After that, the heat capacity of one sample (preferably the smaller one) can be used to extract the additive contribution from the addenda (see Fig. 4). The mass of the thermocouple cross made of a pair of 2 mm long and 25 μm diameter wires is 16 μg, and can be further reduced using thinner wires. The part of the cross which is glued to the sample can be considered as a sample platform, the specific heat of which enters the total heat capacity. For our samples with a surface of 0.5 x 0.5 mm$^2$ this represents ~4 μg of thermocouple material. The remaining part represents the heat link. Assuming that ~1/3 of the heat link contributes to the addenda for $\omega\tau \approx 1$[5], we find that ~8 μg of the thermocouple material should enter the total heat capacity. By using literature data[20] to estimate the heat capacity of the thermocouple, we find that it makes up at least 95 % of the addenda at 100 K (i.e. 17 % of the total heat capacity in the case of the 48 μg Ag calibration sample). This is not a serious limitation since it can be determined with sufficient accuracy as explained. The remaining part of the addenda comes from the glue and remains small (< 1 % of the total heat capacity). This is due to the high dilution factor of the varnish. Based on literature data for the specific heat of GE7031[14], we estimate that the dried mass of varnish deposited onto the sample is ~0.3 μg. Using a syringe fitted with a 0.5 mm diameter needle allows us to obtain a ~30 % reproducibility. Therefore the uncertainty associated with the contribution of the varnish does not exceed ~ 0.5 % for the 48 μg Ag calibration sample. We believe that the addenda contribution can be controlled within ~ 2 % for samples comparable in size with the calibration samples. Adding the uncertainty arising from the conductivity of the heat link, we find that absolute values can be expected within at least ~5 % using samples similar to the calibration samples. This is comparable to the accuracy found for modulated-bath calorimeters of a different design[1]. For very small samples (a few μg's) the uncertainty in the absolute value nevertheless becomes important, but the resolution remains high, so that the study of phase transitions becomes the major field of application of such a calorimeter.

The sensitivity of the thermocouples is usually field dependent, but this cancels out in the formulas of the heat capacity, as only ratios of the signals of thermocouples of the same kind enter. There remains the possibility that the heat conductance $k$ of the thermocouple cross changes with the field. This effect was found to be negligible by measuring the heat capacity of an Ag sample (which is not field dependent) in a slow temperature sweep upon changing the field stepwise from 0 to 14 T. No anomalies due to the field changes appeared in the final specific-heat data, thus setting an upper limit of ~1 % to field-induced errors.

## VI. EXAMPLES OF MEASUREMENTS, RESOLUTION AND ACCURACY

As an illustrative example of the performance of our device, we have chosen a single crystal of the high-temperature superconductor $Bi_{2.12}Sr_{1.71}Ca_{1.22}Cu_{1.95}O_y$ (Bi-2212) close to optimal doping with a mass of 26 μg. Details of sample growth and heat treatment can be found elsewhere[21,22]. Fig. 5a shows the temperature dependence of the amplitude of the thermocouple signals $\Delta T_{S\text{-}B}$ and $\Delta T_{P\text{-}B}$, allowing us to extract the specific heat from Eq. (13). Once the thermal conductance of the heat link and the contribution of the addenda is known, the specific heat can be extracted from either the phase or the amplitude of the thermocouple response. As our DC amplifiers add a frequency-dependent phase shift, we found it more convenient to use the amplitude measurement which also appeared to be less noisy. We have chosen to work at the fixed frequency of $\nu = 0.89$ Hz, nearly 5 times the cutoff frequency at 90 K. In a first step, we tested that this frequency fulfills the condition $\omega > \omega_0$ over the whole temperature range. Fig. 5b shows the corresponding temperature dependence of the of the amplitude of the thermocouple signals $\Delta T_{S\text{-}P}$ and $\Delta T_{P\text{-}B}$ used in the modulated-bath mode to calculate the specific heat using Eq. (7). Here the working frequency was 0.1 Hz, about half the cutoff frequency at 90 K. We selected the standard AC-calorimetry mode, using thermocouple $\Delta T_{S\text{-}B}$, as this allows us to work at a higher frequency and thus to sweep the temperature faster. The working frequency of 0.89 Hz was low enough to use DC preamplifiers. To obtain maximum resolution, we smoothed the temperature dependence $|\Delta T_{P\text{-}B}(T)|$, which does not contain any structure, before calculating $|\Delta T_{P\text{-}B}(T)| / |\Delta T_{S\text{-}B}(T)|$. The result in fields from zero to 14 T is presented in Fig. 6a. The absolute value of the specific-heat curves agrees within 1.4 % at 100 K (see Fig. 6b) with previously published specific heat data obtained on a ~10'000 times larger single crystal of Bi-2212 (0.2545 g) using an adiabatic calorimeter[21].

We note that the size of the anomaly at the superconducting transition of Bi-2212 is very small compared to the large phonon background. Using the difference of specific heat in zero field and in 14 T (see insert in Fig. 6) to estimate the size of the specific-heat anomaly at the transition temperature $T_c$, we find that the anomaly is only ~0.5 % of the total specific-heat value, which can still be resolved with good resolution in this type of calorimeter. For the measurements we set the time constant of the lock-in amplifier to 1 s (i.e. the signal is practically not averaged) and recorded data during a temperature sweep of ~5 mK/s. To characterize the noise in the measurement we subtracted a smooth polynomial from the structureless 14 T data of Fig. 6. The noise level as characterized by the root mean square is 2.6 x $10^{-5}$ J $K^{-1}$ $gat^{-1}$. The limit of resolution due to noise is thus 0.02 % of the total specific heat at $T$ = 90 K (see Fig. 7), 30 times smaller than the anomaly at $T_c$.

## VII. CONCLUDING REMARKS

In the following we would like to compare the performance of our device to other types of calorimeters. The measurements show that this design allows us to measure the specific heat of samples of tiny masses of a few tens of micrograms, or even down to a few micrograms, with a sensitivity being comparable to standard adiabatic methods. Clearly, when the sample mass decreases, the absolute accuracy becomes increasingly sensitive to uncertainties in the addenda contributions and in the heat conductance of the thermal link. Best absolute values are obtained with adiabatic methods, but require samples with masses ~0.1 to 10 g[19,23,24]. Conventional relaxation methods[4,25-27] can be used for samples with masses down to ~100 mg, the limit of miniaturization being given by the fact that a thermometer and thus a sample platform are needed. This limit has been pushed down to a few mg's by the use of $Si_3N_4$ membranes onto which thin-film heater, thermometers and electrical leads were deposited[8,28]. In AC methods[18], thermocouples can be used to measure the sample temperature relative to a heat bath. The addenda can be reduced to a minimum by using chopped light from a laser diode[6-7,9] or a modulated-bath technique[1,2] to heat the sample. This feature, the high sensitivity due to signal averaging and the ease of construction make it an excellent choice for miniature calorimeters with $10^{-4}$ to $10^{-5}$ relative accuracy. One advantage of AC calorimetry lies in the possibility to measure the field dependence of the specific heat during a field sweep at stabilized temperature with a high density of data.

The calorimeter presented in this article is specialized for use in the temperature range between 30 to 300 K and for sample masses from a few mg's up to a few hundreds of mg's, when the main interest is to study phase transitions with a high resolution. Like most AC methods, it is less well suited for the determination of accurate absolute values. Compared to other miniature calorimeters[8,28] an important advantage of the present design is that no equipment and know-how associated with clean rooms, thin-film deposition, photolithography, etc., is needed. Essentially this method only requires a $10 Peltier element. The lower temperature limit is set by the fact that the Peltier power decreases faster than linearly when $T \to 0$. A similar design, with a modulated bath rather based on Joule heating in combination with AuFe/Chromel thermocouples, may be used at lower temperatures, but the advantage of having no DC offset in the measured signal is lost.

It should be remembered that AC (and relaxation) calorimetry may underestimate latent heats at first-order transitions when the amplitude of the temperature modulation is smaller than the temperature hysteresis. Finally note that larger and smaller Peltier elements are commercially available[10], which can be used in combination with different thermocouple wires to construct calorimeters best suited to the samples under study. Two such elements are mounted in the device shown in Fig. 1c.

## ACKNOWLEDGMENTS

We would like to thank B. Revaz for providing the Bi-2212 sample and T. Plackowski for fruitful discussions. This work was supported by the Swiss National Science Foundation through the National Centre of Competence in Research “Materials with Novel Electronic Properties-MaNEP”.

## Figure captions

FIG. 1. (a) Schematic diagram of the Peltier element formed of n- and p-type Bi-Te alloys used to modulate the sample temperature $T_S$. The sinusoidal voltage $U_{ac}$ from the output of a lock-in amplifier[17] is fed into the Peltier element. (b) Schematic diagram of the Peltier AC calorimeter (view from top). The voltages read on three type E Chromel-Constantan thermocouples measure the following temperature differences: $U_{S\text{-}P}$ between the sample and the top plate of the Peltier cell, $U_{S\text{-}B}$ between the sample and the heat sink (or main bath), and $U_{P\text{-}B}$ between the top plate of the Peltier cell and the main bath (see text). (c) Platform holding two independent calorimeters with their Peltier cells. The smaller one, 3.2 x 3.2 mm (as used for the measurements presented in this article), uses 25 μm ∅ thermocouples, whereas the larger one, 6.4 x 6.4 mm, uses 50 μm ∅ thermocouples.

FIG. 2. Schematic diagram of the calorimeter, where the shaded area represents the top plate of the Peltier element which can be considered as a modulated thermal bath of temperature $T_P(t)$, $T_S(t)$ is the sample temperature and $T_B$ the temperature of the main thermal bath. The heat conductance $k$ is determined by the legs of the thermocouple cross between the sample and the point where they are thermally anchored to the top plate of the Peltier element. $k_P$ is the heat conductance between the top and base plate of the Peltier element. The double-sided arrows show the temperature differences $\Delta T_{S\text{-}P}$, $\Delta T_{P\text{-}B}$ and $\Delta T_{S\text{-}B}$ as measured by the three thermocouples (see Fig. 1).

FIG. 3. (a) Frequency dependence of the amplitude of the thermocouples measuring $\Delta T_{S\text{-}B}$, $\Delta T_{S\text{-}P}$ and $\Delta T_{P\text{-}B}$ for a 26 μg $Bi_{2.12}Sr_{1.71}Ca_{1.22}Cu_{1.95}O_y$ ('Bi-2212') sample at 90 K (measured without preamplifier or transformer). Taking into account phase shifts, the complex relation $\Delta T_{S\text{-}B} = \Delta T_{S\text{-}P} + \Delta T_{P\text{-}B}$ is obeyed. (b) Frequency dependence of the amplitudes $|\Delta T_{S\text{-}B}|$ and $|\Delta T_{S\text{-}P}|$ divided by the excitation $|\Delta T_{P\text{-}B}|$. The lines are one-parameter fits using Eq. (5) and (11). The crossing point of the two lines represents the cutoff frequency $\nu$=0.18 Hz. The data at high frequencies of the thermocouple $\Delta T_{S\text{-}P}$ fall below the fit, possibly due to an asymmetry of heat flow through the $\Delta T_{S\text{-}B}$ and $\Delta T_{S\text{-}P}$ thermocouples. The data taken above 4 Hz are excluded from the analysis (see text for details). (c) Dependence of the amplitudes $|\Delta T_{S\text{-}B}|$, $|\Delta T_{S\text{-}P}|$ and $|\Delta T_{P\text{-}B}|$ on the voltage $U_{ac}$ (effective value) driving the temperature modulation of the top plate of the Peltier element. In this test conducted at $T$ = 90 K and $\nu$ = 0.89 Hz, a 26 μg Bi-2212 sample is attached to the thermocouple cross.

FIG. 4. Total heat capacity of a pure (99.999 %) Ag sample (48 μg) including the addenda (top curve) compared to reference data for the sample alone (dotted line)[19]. The difference allows one to estimate the addenda heat capacity of the thermocouple cross and the glue (GE7031[13]) used to mount the sample (see text for details).

FIG. 5. (a) Temperature dependence of the amplitude $|\Delta T_{S\text{-}P}|$ and $|\Delta T_{P\text{-}B}|$ of the thermocouple signals. The data are taken at $\nu$=0.1 Hz, slightly below the cutoff frequency, to be analyzed in the modulated-bath scheme. (b) Temperature dependence of the amplitude $|\Delta T_{S\text{-}B}|$ and $|\Delta T_{P\text{-}B}|$ of the thermocouple signals. Here the data are taken at $\nu$=0.89 Hz, about five times the cutoff frequency, to be analyzed within the AC method.

FIG. 6. a) Specific heat *C/T* of a single crystal of Bi-2212 with a mass of 26 mg in fields of 0 T, 1 T, 2 T, 8 T and 14 T measured with the present modulated-bath calorimeter. The data are taken at n=0.89 Hz and analyzed within the AC method. One gram-atom is 59.4 g (1/15 mole). The anomaly at 90 K is the superconducting transition. In the insert a background given by the smoothed 14 T data has been subtracted. b) Zero-field specific-heat data *C* of the 26 mg Bi-2212 sample compared to literature data[21].

FIG. 7. Residual noise in the specific-heat data after subtracting a smooth polynomial from the 14 T data in Fig. 6. The root-mean-square value of the scatter is 2.6 x $10^{-5}$ J $K^{-2}$ $gat^{-1}$ or 0.023 % of the sample specific-heat.

(a)

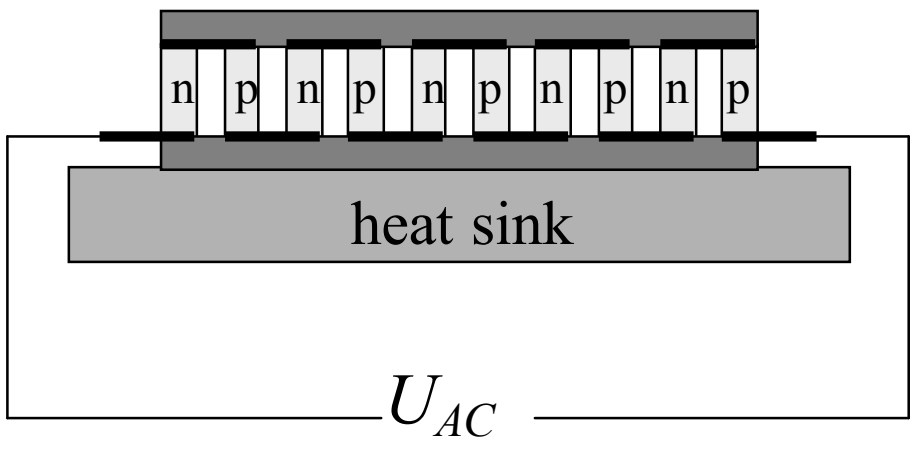

(b)

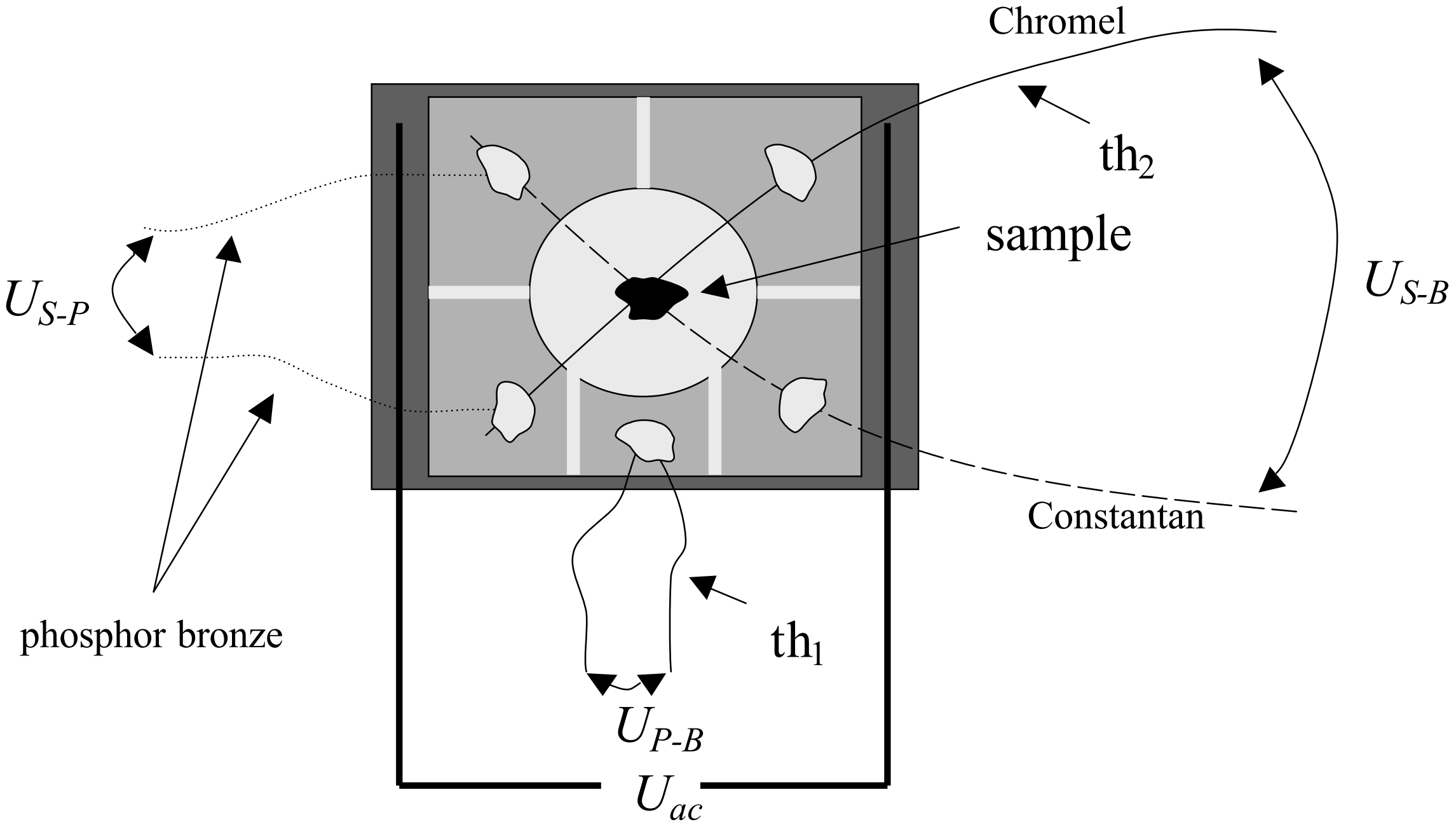

(c)

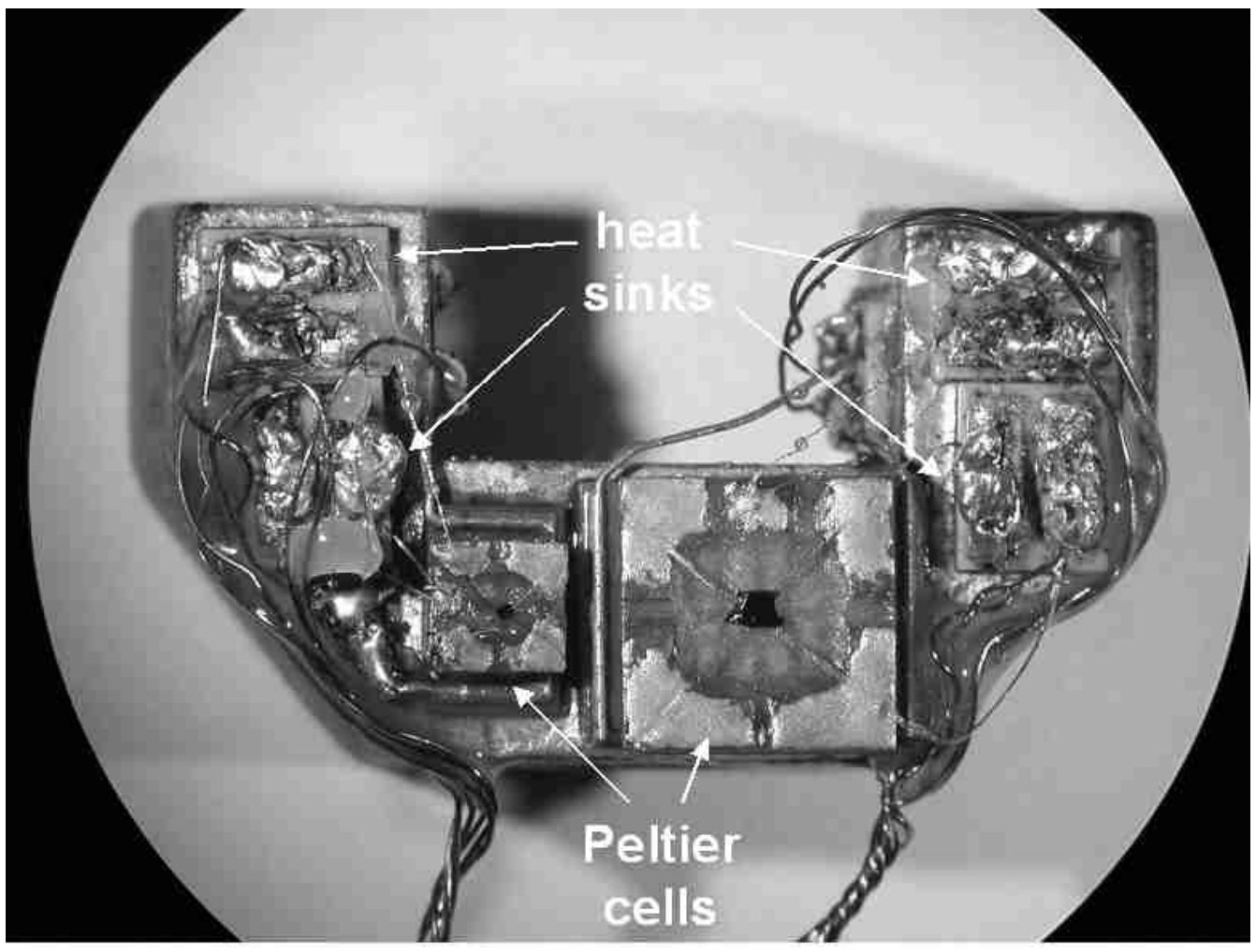

Fig. 1a
Fig. 1b
Fig. 1c

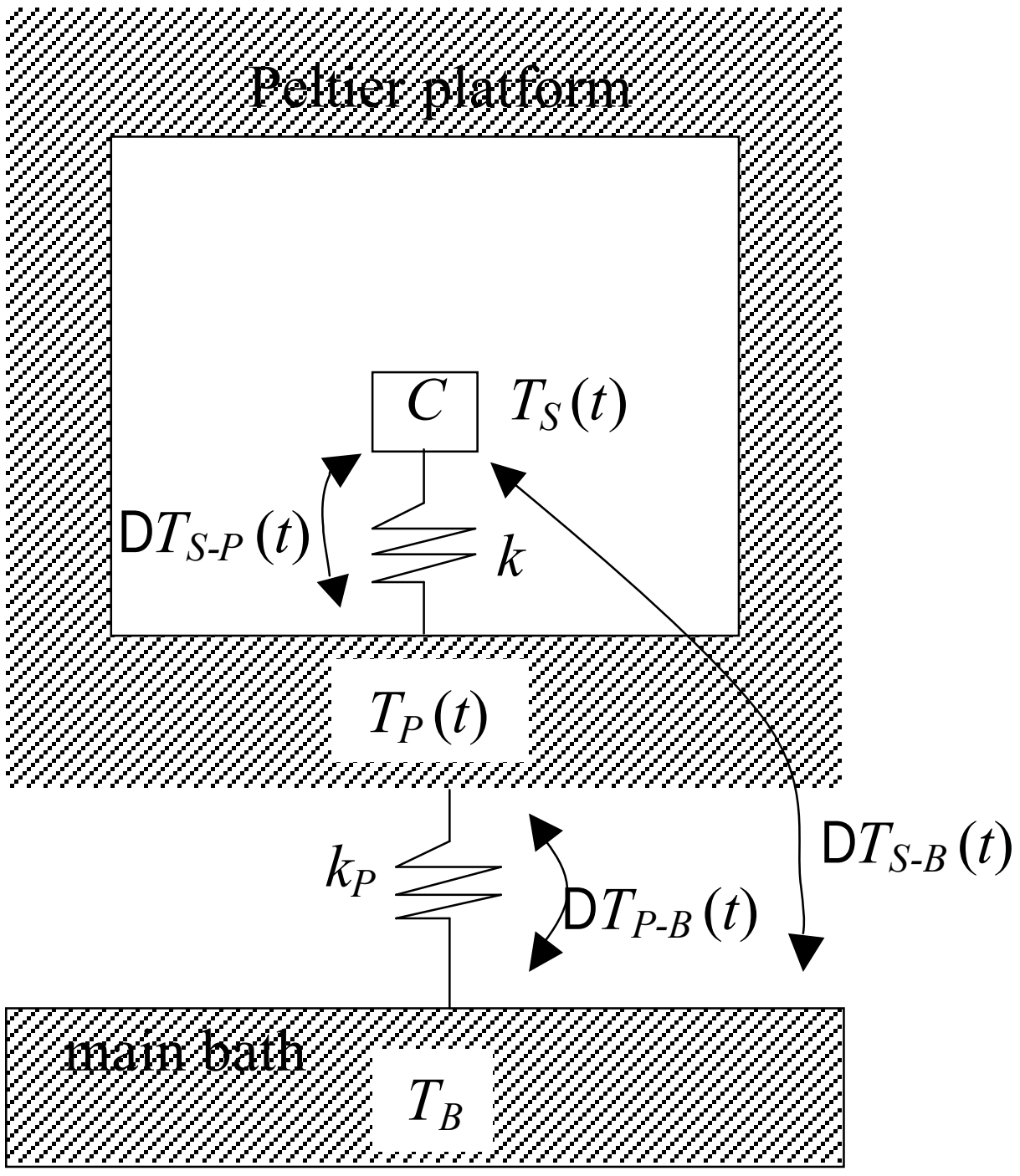

Fig. 2

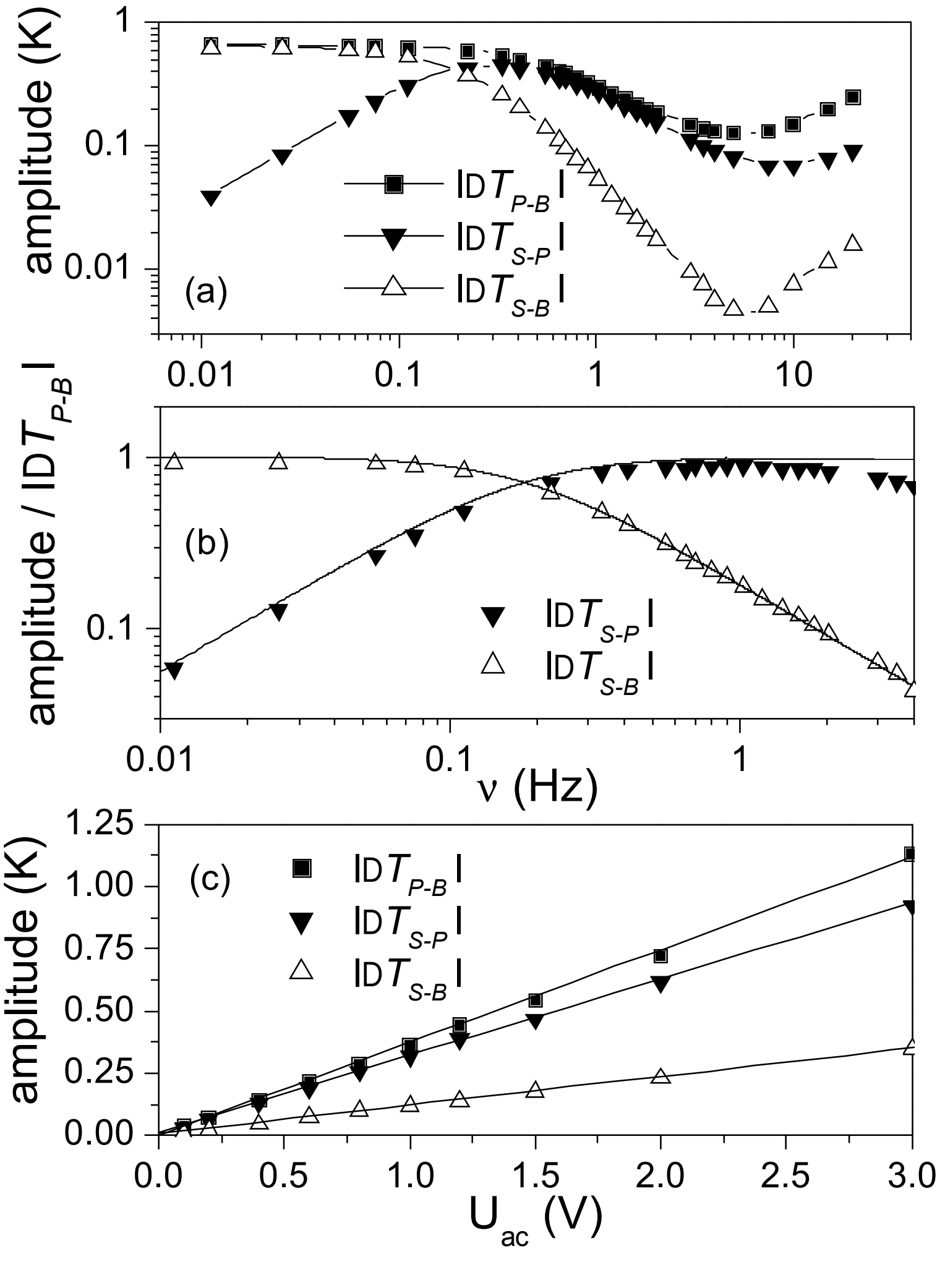

Fig. 3a
Fig. 3b
Fig. 3c

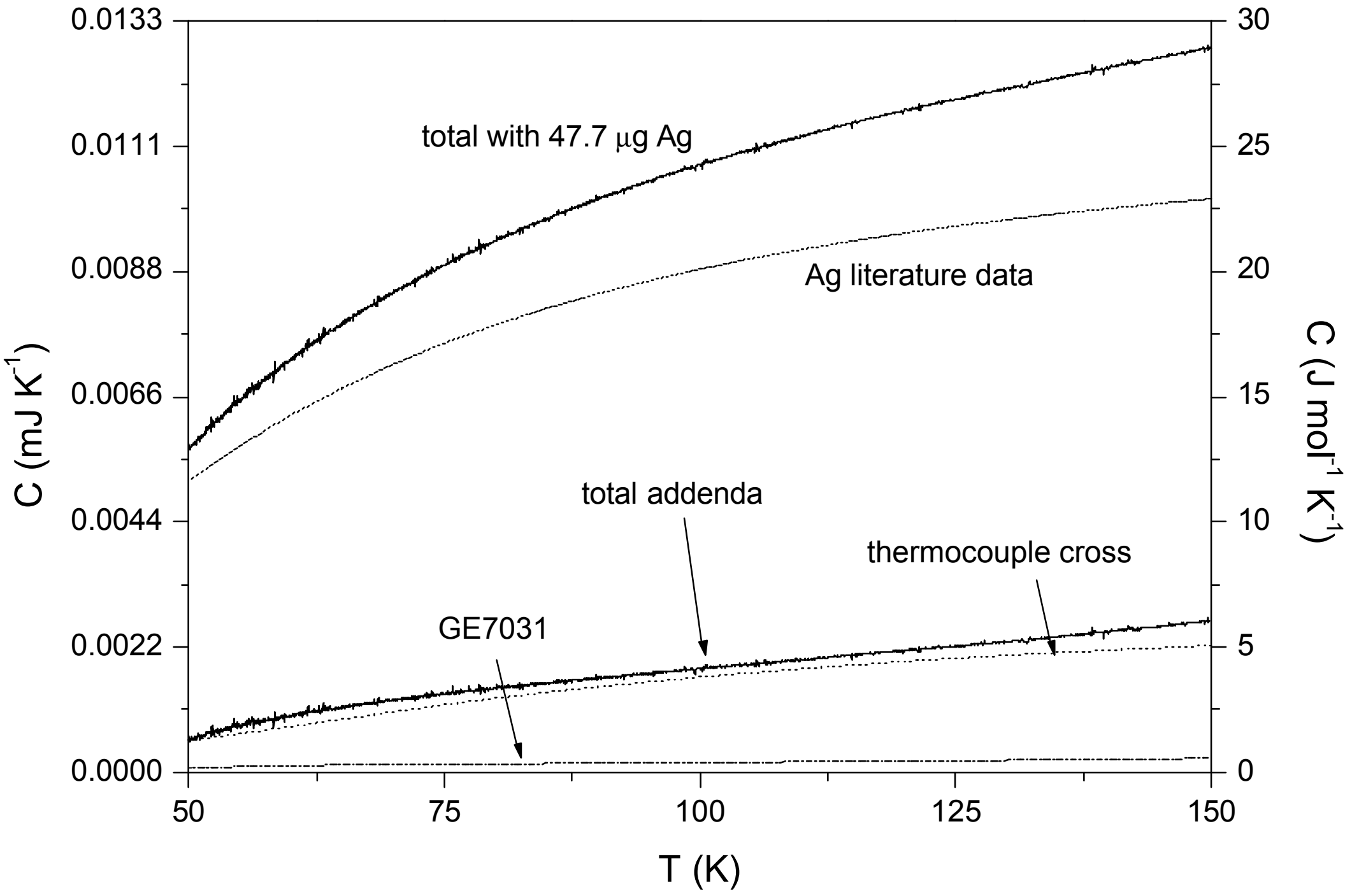

Fig. 4

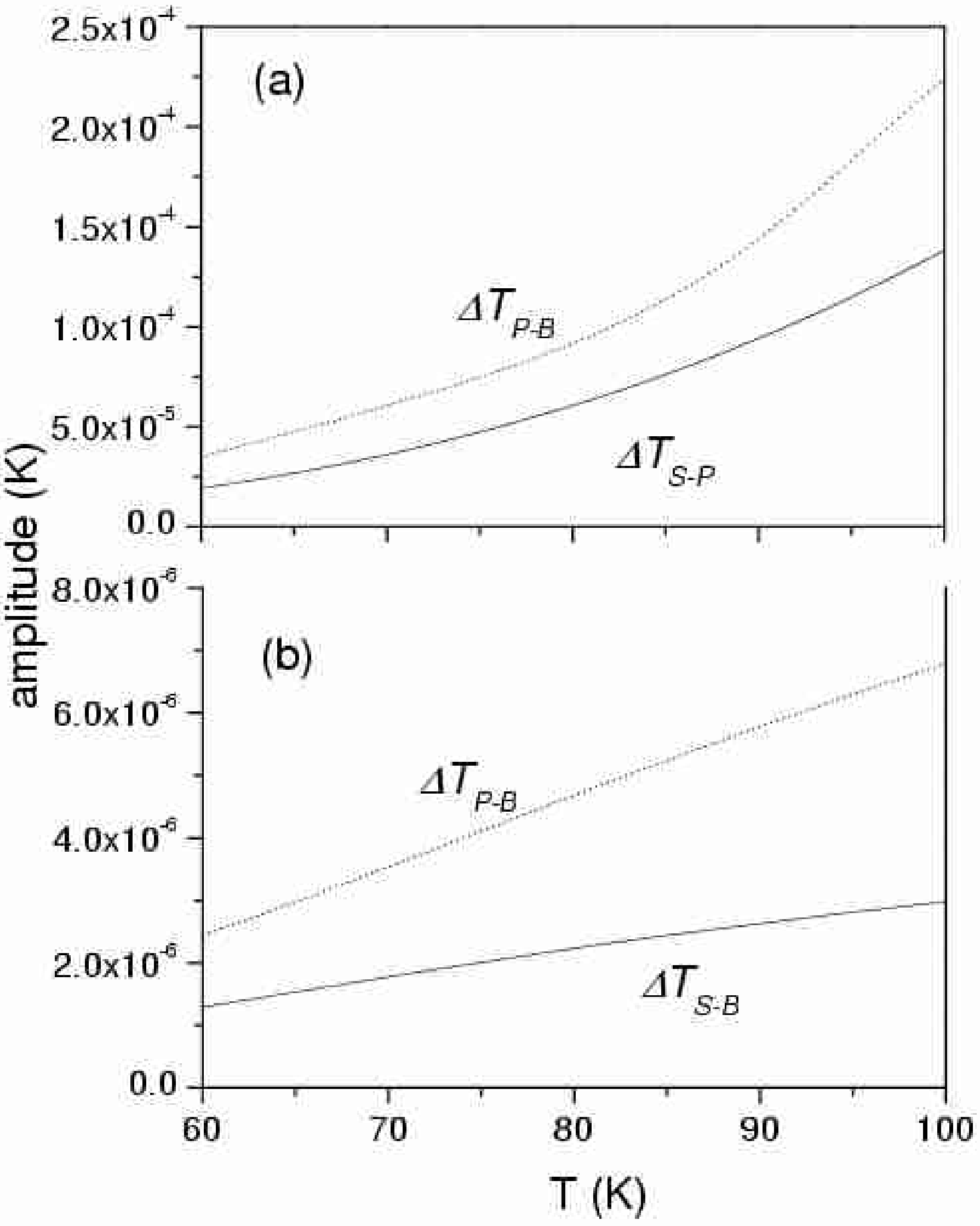

Fig. 5a
Fig. 5b

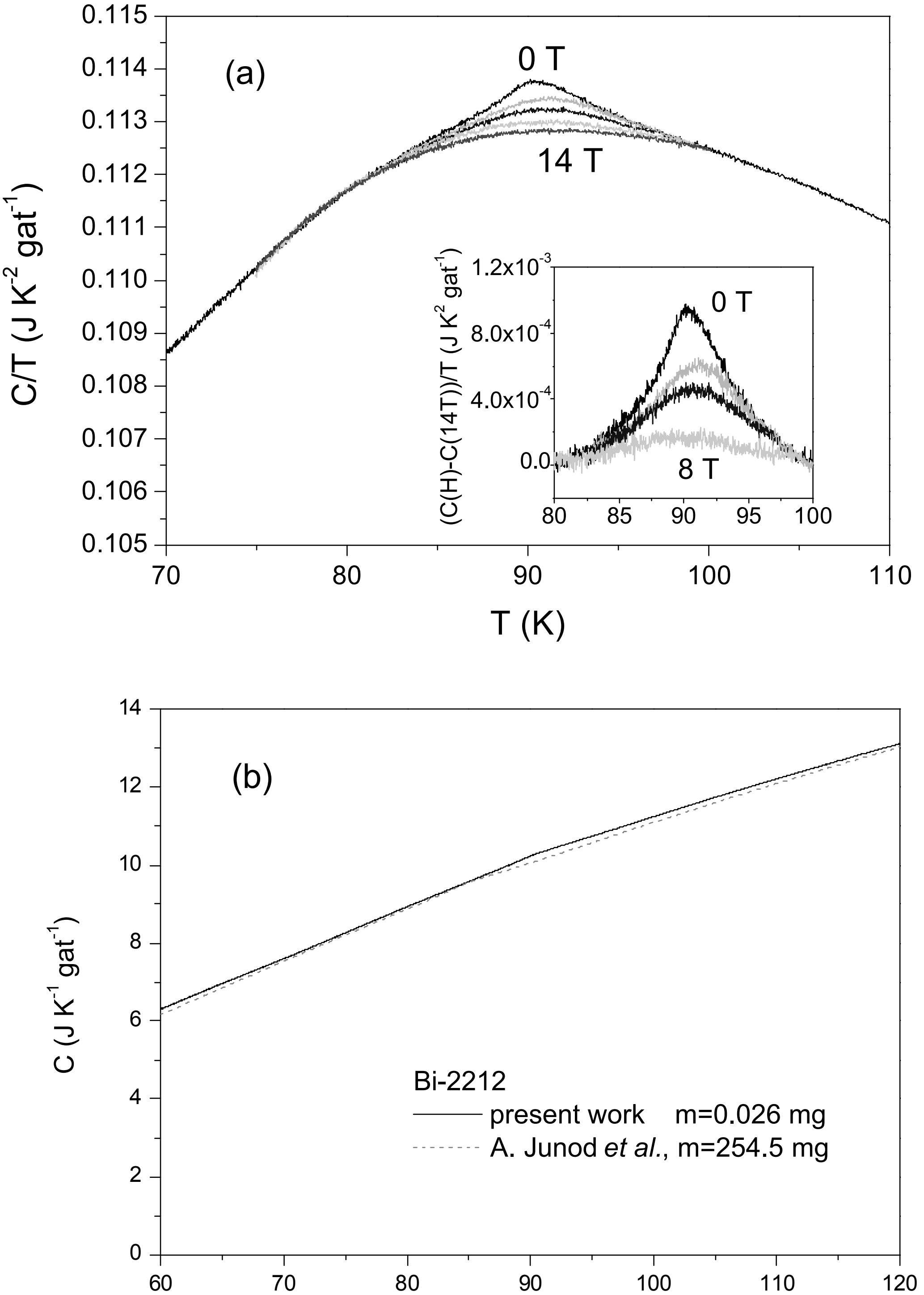

Fig. 6a
Fig. 6b

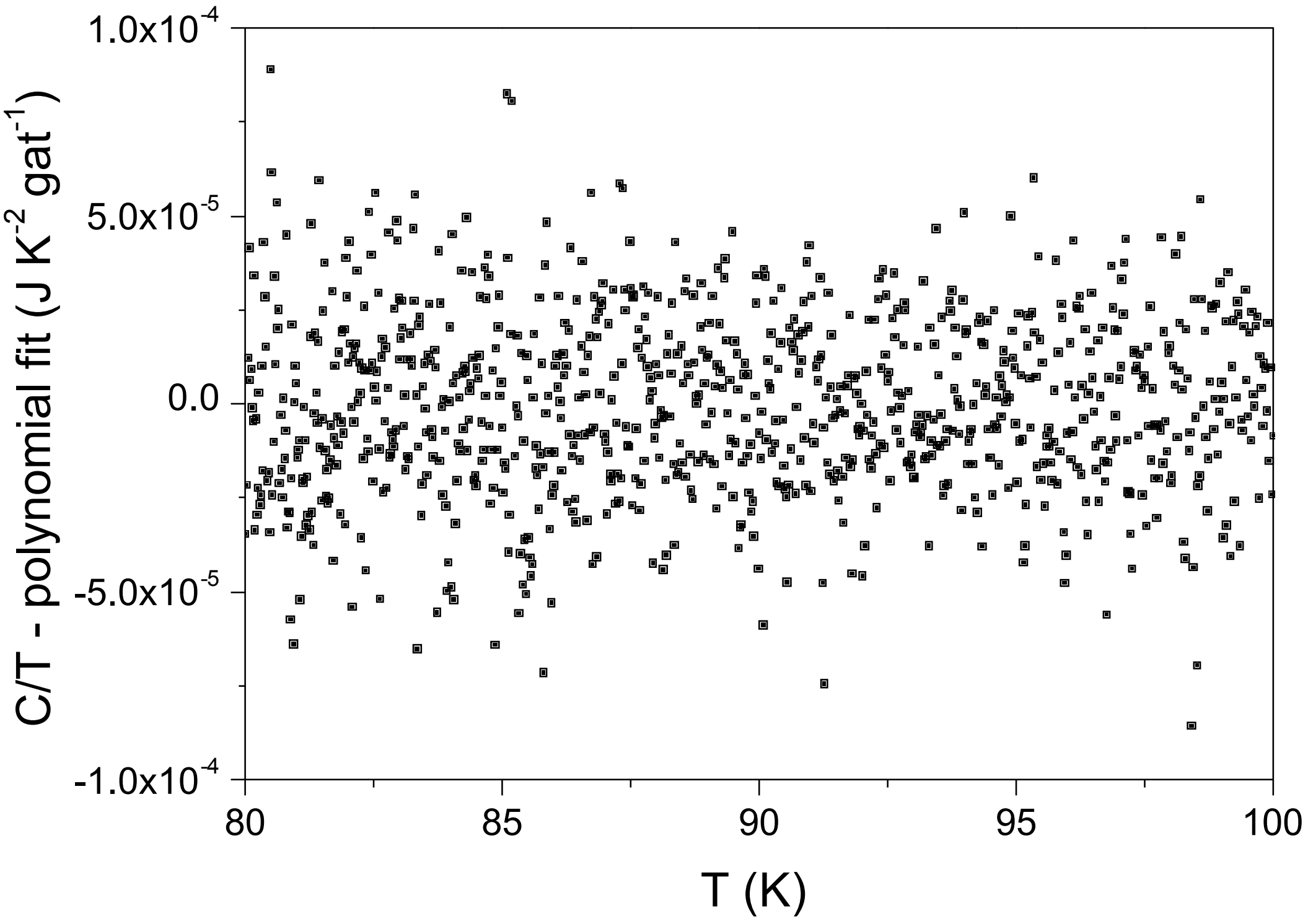

Fig. 7